\documentclass[aps,notitlepage,twocolumn,nofootinbib,pra,10pt,usenames]{revtex4-2}
\usepackage{newtxtext,amsmath,amsfonts,amssymb,amsthm,bbm,enumerate,times,mathtools,physics,marvosym,wasysym,soul,lmodern,stmaryrd,dsfont,graphicx}
\usepackage[usenames,dvipsnames]{xcolor}
\usepackage{float}
\usepackage{hyperref}
\hypersetup{colorlinks,linkcolor=black,citecolor=ForestGreen,urlcolor=black}
\usepackage[ruled,longend]{algorithm2e}
\usepackage{etoolbox}
\usepackage{dsfont}
\usepackage[noabbrev,capitalise]{cleveref}
\usepackage{comment}
\usepackage{xfrac}
\DeclareMathOperator*{\argmax}{arg\,max}
\usepackage{multirow,makecell}

\pdfoutput=1
\usepackage[inline]{enumitem}
\newcommand{\codepar}[1]{\llbracket #1 \rrbracket}

\begin{document}

\title{Biased-Noise Thresholds of Zero-Rate Holographic Codes with Tensor-Network Decoding}
\author{J.~Fan$^{1,2\,\dagger}$, M.~Steinberg$^{1,2\,\dagger,\ddagger}$, A.~Jahn$^{3}$, C.~Cao$^{4}$, S.~Feld$^{1,2}$}
\affiliation{$^{1}$QuTech, Delft University of Technology, 2628 CJ Delft, The Netherlands}
\affiliation{$^{2}$Quantum and Computer Engineering Department, Delft University of Technology, 2628 CD Delft, The Netherlands}
\affiliation{$^{3}$Department of Physics, Freie Universit\"at Berlin, 14195 Berlin, Germany}
\affiliation{$^{4}$Department of Physics, Virginia Tech, Blacksburg, VA 24061, USA}

\begin{abstract}
A crucial insight for practical quantum error correction is that different types of errors, such as single-qubit Pauli operators, typically occur with different probabilities. Finding an optimal quantum code under such biased noise is a challenging problem, related to the (generally unknown) maximum capacity of the corresponding noisy channel. A benchmark for this capacity is given by the hashing bound, which describes the performance of random stabilizer codes and leads to the matter of identifying codes that come close to the bound while also being efficiently decodable. In this work, we perform the first comprehensive analysis of asymptotically zero-rate holographic codes under biased noise. We show that many representatives from such models of this code class fulfill both the channel optimality and efficient decoding guarantees for tensor-network codes. In fact, all holographic codes tested were found to reach the hashing bound in some bias regime, while several built from the $\codepar{5,1,2}$ surface code and $\codepar{6,1,3}$ code exceed state-of-the-art code performance in the 2-Pauli noise regime. Furthermore, we consider Clifford deformations which allow all considered codes to reach the hashing bound for 1-Pauli noise as well. Our results establish that holographic codes, which were previously shown to possess efficient tensor-network decoders, also exhibit competitive thresholds under biased noise.
\end{abstract}

\maketitle

\begingroup
  \renewcommand{\thefootnote}{\fnsymbol{footnote}}
  \footnotetext[2]{These authors contributed equally to this work.} 
  \footnotetext[3]{Corresponding author: matt.steinberg3@gmail.com} 
\endgroup

\section{Introduction}\label{section:intro}

Holographic codes are quantum codes conventionally studied as toy models of the Anti-de Sitter/Conformal Field Theory (AdS/CFT) correspondence~\cite{witten, maldacena, klebanov,almheiri2015bulk,harlow2017ryutakayanagi,happy_paper,Harlow:2018fse,holographic_codes_topical_review} where they have led to remarkable insights in quantum gravity. However, because their relevance in quantum gravity is mostly theoretical~\cite{uberholography,harlow2017ryutakayanagi}, very little is known about their practical error correction capabilities in contrast to leading benchmarks of quantum error correction, such as the surface code. In addition, recent advances in tensor network codes~\cite{farrelly_tn_codes} and the quantum LEGO formalism~\cite{cao_quantum_lego} have brought them closer to a practically viable regime. For example, leveraging their flexibility in the modular construction, holographic codes with high encoding rate and greatly exceeding the $O(n^{1/2})$ distance scaling have been proposed~\cite{evenbly_codes2} alongside schemes for implementing universal fault-tolerant logic based on heterogeneous code constructions~\cite{heterogeneous_holo_qrm}. The hyperbolic geometry of holographic codes also implies that such codes admit efficient optimal tensor network decoders for all i.i.d. single qubit errors~\cite{cao2024quantum} since their tensor networks can be contracted in polynomial time even without bond truncation. Under such decoders, they have been shown to yield a code capacity threshold that are as high as that of the surface code~\cite{farrelly_tn_codes} under depolarizing noise where all Pauli errors occur with equal probability. 

Although there remain many open questions before holographic codes can be fairly assessed at a practical level, e.g. determining the fault-tolerant threshold, there are properties that are poorly understood even at the code capacity level, where syndrome extractions are perfect. One such open question is the performance of the code when the error channels are biased. Biased noise is found in many physical quantum devices wherein certain Pauli errors such as phase flips, occur more frequently than bit flips. The presence of bias can also lead to better error correction with superior threshold and effective distances~\cite{sahay2023high,xzzx_surface_code,tailored_513_xzzx_codes,clifford_deformed_surface}. In certain topological codes (e.g. Clifford-deformed topological codes such as the XZZX surface code) the threshold under biased noise has also been shown to attain the error threshold set by the hashing limit, which sets a target benchmark for good quantum error correcting codes that can be used for efficient information transmission under a noisy channel. In holographic codes, however, there is very little understanding of its performance against biased error~\cite{cao2024quantum,parallel_decoding_tn_codes}.  

In this work, we provide the first characterization of holographic codes under biased noise. We study common classes of holographic codes and their deformations to demonstrate that asymptotically zero-rate \emph{holographic quantum codes} admit high code capacity thresholds comparable to leading benchmarks under biased noise and can either achieve or supersede the hashing bound for various Pauli channels of interest. The possibility of holographic codes as an alternative class of code constructions to exceed the hashing bound is also of independent interest, as a central problem in quantum coding theory is to identify efficiently decodable codes that can attain or even exceed the hashing bound. A class of such codes will not only provide new approaches for obtaining tighter \emph{quantum channel capacity} (QCC) lower bounds but also inspire proposals for more efficient strategies of error correction~\cite{qccapacity_very_noisy,degenerate_qu_code_pauli_channels,lower_bounds_capacity_birgitta,leditzky_pauli_hashing,xzzx_surface_code}.

Using tensor-network decoding techniques from~\cite{farrelly_tn_codes,parallel_decoding_tn_codes}, we study zero-rate versions of six different classes of holographic codes which are constructed from the seed codes of the 
\begin{enumerate}[label=(\alph*), itemjoin={{; }}, itemjoin*={{; and }}, nosep]
  \item $\codepar{5,1,3}$, i.e. Harlow–Preskill–Pastawski–Yoshida (HaPPY) code~\cite{happy_paper}
  \item Tailored $\codepar{7,1,3}$ code~\cite{tailored_713}
  \item $\codepar{7,1,3}$ Steane code~\cite{harris_css}
  \item $\codepar{6,1,3}$ code~\cite{farrelly_tn_codes}
  \item Surface-code fragment (SCF)~\cite{harris_iod}, and
  \item A Clifford-deformed variant of the Steane code.
\end{enumerate}
Our main results can be stated as follows. In the limit of pure 2-Pauli noise channels, the thresholds calculated for zero-rate SCF and $\codepar{6,1,3}$ holographic codes exceed current state-of-the-art results~\cite{leditzky_pauli_hashing,lower_bounds_capacity_birgitta,qccapacity_very_noisy}. To our knowledge, no other class of quantum code has thus far been able to supersede the hashing limit for pure 2-Pauli noise. In fact, all of the holographic codes tested in the limit of pure 2-Pauli noise either come within $2\%$ of, attain, or overcome some portion of the hashing bound. Furthermore, all codes tested were found to either attain or overtake the hashing bound for some bias configuration of Pauli noise channel. Finally, we also apply the method of \emph{Clifford deformations} from~\cite{3d_surface_hashing,clifford_deformed_surface} to the holographic Steane code, demonstrating that the threshold of the code can be drastically modified for depolarizing, pure X, and pure Z noise channels. Our results provide strong evidence that holographic codes possess high resilience against various types of Pauli noise channels and can be tuned accordingly. In sharp contrast with other code families, maximum likelihood decoding of holographic codes and the computation of their weight enumerator polynomials are possible in polynomial time, even with exact tensor network contractions~\cite{farrelly_tn_codes,cao2024quantum}, further bolstering their practical potential for both quantum error correction and for numerical studies of QCC bounds.

The remainder of this article is organized as follows: we review in the background (\cref{section:background}) the basics of holographic codes and their properties for practical QEC in \cref{section:background_hqec_codes}, as well as an explanation for how tensor-network decoding functions (\cref{section:TN_decoding}). In \cref{section:results}, we begin by discussing first the experimental setup for our simulations, the codes tested, and the biased-noise points considered (\cref{section:noise_model}); in \cref{section:zero_rate_results}, we present all of the results of our study. Finally, we close with a discussion of future directions and implications of our work (\cref{section:discussion}), followed by a discussion of implications for quantum coding theory (\cref{section:discussion_quantum_coding_theory}), as well as practical directions for holographic codes (\cref{section:discussion_holo_codes_in_practice}). 

\section{Background} \label{section:background}

\subsection{Holographic Quantum Error Correction}\label{section:background_hqec_codes}

Originally developed to model the \emph{AdS/CFT correspondence}~\cite{witten,maldacena,klebanov,nastase,takayanagi_book}, holographic quantum error correction involves encoding maps from \emph{bulk} (logical) degrees of freedom on a $d{+}1$-dimensional hyperbolic space to $d$-dimensional \emph{boundary} (physical) degrees of freedom. While in full AdS/CFT the bulk and boundary degrees of freedom are associated with weakly-coupled quantum gravity on an asymptotically anti-de Sitter (AdS) background and strongly-coupled conformal field theory (CFT), respectively, aspects of this duality for $d=1$ can be captured by simple tensor network codes~\cite{almheiri2015bulk,happy_paper,Harlow:2018fse,holographic_codes_topical_review}.

\begin{figure*}[ht]
\centering
\includegraphics[width=0.9\textwidth]{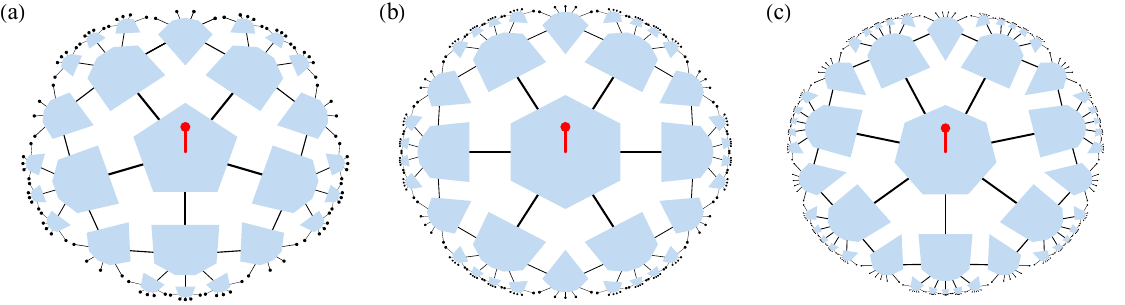}
\caption{The holographic tensor-network codes considered in this work: (a) The \emph{Harlow-Preskill-Pastawski-Yoshida} (HaPPY) code~\cite{happy_paper} whose pentagon-hexagon geometry also underlies the holographic \emph{surface-code fragment} (SCF) model~\cite{harris_iod}, (b) the holographic $\codepar{6,1,3}$ code~\cite{farrelly_tn_codes}, and (c) the holographic Steane code~\cite{harris_css}.
All are defined as tensor network contractions of copies of a fixed $q$-leg tensor on a hyperbolic tiling, with the central tensor used as a \emph{seed tensor} for the encoding isometry of a $\codepar{q-1,1,d}$ code with some distance $d$.
The remaining tensors have all $q$ legs contracted in the plane, leading to a larger encoding map between one logical qubit (central red leg) and the boundary physical qubits (open black legs).
Here we depict the codes with $R=2$ layers of edge inflation. In the $R\to\infty$ limit, the rate of each code goes to zero.}
\label{fig:zero_rate_codes}
\end{figure*}

Holographic quantum codes are, at their core, \emph{generalized concatenated} codes~\cite{poulin2005subsystem,lidar_qec,williamson_ft_logical_gates,gottesman_thesis_stabilizer_codes,heterogeneous_holo_qrm, baspin2024wire, liu2024subsystem}, as their tree-like concatenation structure is a consequence of the bulk logical qubits being divisible into radial layers~\cite{happy_paper,conformal_quasi_holo,central_charges_jahn_eisert}.
The most widely studied instances of holographic codes also fall into the category of \emph{stabilizer codes}~\cite{happy_paper,harris_css,steinberg_evenbly_codes1}, despite some restrictions on their representation of holographic dualities~\cite{Dong:2018seb,Akers:2018fow,Cao:2023mzo} and the possibility to construct more sophisticated non-stabilizer or even approximate holographic codes~\cite{hmera_approx,pollack2022understanding,cao2021approximate,kim2017entanglement,xp_stabilizer}. We shall therefore limit ourselves to stabilizer versions of holographic codes in the present work. 

Stabilizer codes themselves are defined as codes for which the \emph{logical operators} $\mathsf{L}$ and \emph{codeword stabilizers} $\mathsf{S}$ consist of elements $P_{i} \in \mathsf{P}^{n}$, where $\mathsf{P}^{n}$ is the $n$-qubit Pauli group, $P_{i}$ takes the form $P_{i_{1}} \otimes \cdots \otimes P_{i_{n}}$, and $P \in \{ I, X, Y, Z\}$, i.e., the single-qubit Pauli operators. It is also wholly possible to consider \emph{finite-rate} versions of holographic codes, wherein individual seed tensors can be specified with logical-qubit implantation. In this work, we consider only zero-rate versions of holographic codes, wherein only the central logical index remains (\cref{fig:zero_rate_codes})\begingroup
\renewcommand{\thefootnote}{\fnsymbol{footnote}}
\footnote[1]{One can additionally construct subsystem versions of holographic quantum codes. Here, one chooses to subdivide the physical Hilbert space as $\mathsf{H}_{p} = \mathsf{H} \otimes \mathsf{\bar{H}}$, where $\mathsf{H}$ represents the code subspace. Then, given $\mathsf{H}$, we can decompose the code subspace further as $\mathsf{H} = \mathsf{H}_{L} \otimes \mathsf{H}_{G}$, where $\mathsf{H}_{L}$ and $\mathsf{H}_{G}$ represent the \emph{logical} and \emph{gauge} subsystems of the code subspace, respectively. Holographic codes have been previously treated as subsystem stabilizer codes in~\cite{cao2021approximate,williamson_ft_logical_gates,uberholography}.}
\endgroup.

Represented as tensor networks, the bulk geometry of holographic codes is that of a regular hyperbolic tiling, wherein each $n$-gon tile is filled out by one tensor of $n$ planar (i.e., non-logical) legs. Starting from the central seed tensor whose single logical leg describes degrees of freedom in a small bulk patch in holography, the tensor network (and the bulk region it represents) can be iteratively extended via \emph{inflation steps}, Here, each of these extensions can be realized by adding a layer of tensors (contracted) to the tensor network~\cite{conformal_quasi_holo,central_charges_jahn_eisert}. While this process is not unique, we will only apply \emph{edge inflation} in this work (i.e. tensors/tiles are added to all open legs/edges of the previous step). We denote the number of inflation steps as $R$, with $R=0$ (no step) describing the seed tensor by itself, with \cref{fig:zero_rate_codes} showing holographic codes on various hyperbolic tilings at $R=2$.

From a practical perspective, there are many reasons to find the concept of a holographic quantum code appealing. Firstly, holographic codes are relatively simple to construct and scale from small, atomized examples to larger example codes. As a direct consequence of the quantum LEGO formalism~\cite{cao_quantum_lego}, concepts such as code structure and transversal logical operations are straightforward to intuit~\cite{heterogeneous_holo_qrm,cao2025growing}. Secondly, the boundary of holographic codes exhibits a quasiperiodic self-similarity~\cite{conformal_quasi_holo,central_charges_jahn_eisert}, allowing one to rescale the number of physical and logical qubits using local inflation rules that also determine the code's rate and distance scaling~\cite{parallel_decoding_tn_codes,evenbly_codes2}, both of which fare better than other constructions~\cite{campbell2017roadstoward,terhal2015quantum,fowler2012surface,breuckmann_qldpc}. Thirdly, it is known that most seed tensors for holographic codes can be easily mapped to graph states, implying that efficient preparation schemes likely exist~\cite{engineering_holography_graph,raissi_graph1,raissi_graph2,helwig_ame_graph}. Lastly, and perhaps most consequentially, it has been shown that holographic codes exhibit high resilience against various noise channels. Indeed, it was first shown in~\cite{happy_paper} that holographic quantum error correction codes exhibit high thresholds against the quantum erasure channel; subsequently, several works have shown the capability of certain codes constructions to protect against depolarizing and 1-Pauli noise, potentially as well as topological codes~\cite{parallel_decoding_tn_codes,harris_css,harris_iod,farrelly_tn_codes,deconfinement,evenbly_codes2}. To our knowledge, a systematic biased-noise threshold study for even zero-rate holographic codes has not been performed, let alone for their constant-rate versions\begingroup
\renewcommand{\thefootnote}{\fnsymbol{footnote}}
\footnote[1]{\cite{cao2024quantum} discusses biased noise in finite-rate HaPPY pentagon codes, but it only captures the non-detectable error probability for a code of fixed length.}
\endgroup. As such, our work takes the first step towards understanding holographic codes under more generalized noise channels.

The main ingredient of a holographic quantum code is the seed tensor, defining the encoding map for a single tile (in \cref{section:miscellaneous_seed_tensors} we will provide descriptions of several seed codes which may be unfamiliar to the reader, such as the SCF and the tailored $\codepar{7,1,3}$ codes). Here we consider the zero-rate case with a logical qubit only on the central tile; on the remaining tiles all tensor legs are planar, effectively dispersing the logical information towards the tiling boundary. Note that in generic holographic codes, each tile may hold a logical qubit, leading to a nonzero asymptotic rate as the number $R$ of layers is increased. 

Currently, several methods exist by which one may create a holographic code. Firstly, one may employ methods of \emph{gauge fixing} for various logical qubits in the bulk. This step leaves each encoding tensor with only physical legs that act isometrically from one layer to the next. In the second case, one can tile the bulk by surrounding a code with parameters $\codepar{n,k,d}$ by codes with parameters $\codepar{n+1,0,d+1}$. In both cases, the contracted tensor network then forms one large encoding isometry from a single logical qubit to the physical boundary qubits given suitable (e.g.\ \emph{perfect}~\cite{happy_paper}) tensors. Here, we opt for the second case. Thirdly, copies of the same seed tensor are typically used to construct such codes, but one may also employ several different seed tensors in similar spirit to \emph{heterogeneous} concatenation methods for tree-style concatenated codes~\cite{concat1,concat2,heterogeneous_holo_qrm}. We note that all of these techniques it was recently shown that holographic codes are an instance of \emph{generalized code concatenation}~\cite{generalized_code_concat,heterogeneous_holo_qrm}.

\subsection{Seed Tensors Used in this Paper}\label{section:miscellaneous_seed_tensors}

Here we provide information on the seed codes utilized in this paper: the \emph{perfect} $\codepar{5,1,3}$ code~\cite{laflamme1996perfect}; the Steane code~\cite{Steane_1996}; the $\codepar{6,1,3}$ code~\cite{shaw2008encoding}; the surface-code fragment (SCF) from~\cite{harris_iod} and the tailored $\codepar{7,1,3}$ code~\cite{tailored_713}. 

Firstly, The famous \emph{perfect} code of~\cite{laflamme1996perfect} is a $\codepar{5,1,3}$ code with cyclically permuted stabilizer support, as shown in \cref{table:seed_tensors}. The perfect code is an example of generalized XZZX toric codes~\cite{tailored_513_xzzx_codes}, and is known to exhibit high resilience against biased 1-Pauli noise~\cite{xzzx_surface_code,tailored_713}. Equally famous in its own right, the \emph{Steane} code is the prototypical example of a \emph{Calderbank-Shor-Steane} (CSS) code, in which two classical codes are utilized in order to construct a quantum code. The Steane code in particular is a \emph{self-dual} CSS code, meaning that the X-/Z- portions of the stabilizer generators are exactly coinciding. More information on both codes can be found in~\cite{lidar_qec,nielsen_chuang}, and the stabilizers and logical operators for the code can be found in \cref{table:seed_tensors}.

The \emph{surface-code fragment} (SCF) is the smallest surface code, and is known to possess a block-perfect encoding tensor, following the work of~\cite{harris_iod}. The stabilizers and logical operators for the SCF seed tensor are shown in \cref{table:seed_tensors}. The block-perfect condition is taken to mean that, given any permutation of cyclic permutations for the SCF tensor indices $\{1, 2, 3, 4, \bar{6}, 5\}$ (where $\bar{6}$ denotes the logical index), we can construct the condition whereby 
\begin{equation}
\text{Tr}_{A}\big[ \ket{\psi}\bra{\psi}\big] \propto \mathbb{I}~,
\end{equation}
where $|A| = \lfloor \frac{n}{2}\rfloor$. The SCF seed tensors are an instance of \emph{planar maximally-entangled} (PME) states~\cite{doroudiani2020planar}, meaning that only adjacent groupings into bipartitions can reduce to maximally mixed states.

The \emph{tailored} $\codepar{7,1,3}$ code is a generalization of the celebrated \emph{perfect} code~\cite{laflamme1996perfect} and was first introduced in~\cite{tailored_713}. The stabilizers and logical operators for the code are listed in \cref{table:seed_tensors}. As stated in~\cite{tailored_513_xzzx_codes,wen2003quantum}, both the perfect code and the tailored $\codepar{7,1,3}$ codes can be regarded as the smallest toric codes with so-called \emph{shifted} boundary conditions. The tailored $\codepar{7,1,3}$ code was shown in~\cite{tailored_713} to exhibit better threshold performance than the Steane code in the setting of highly-biased noise.

\subsection{Tensor-Network Decoding} \label{section:TN_decoding}

Contracting a tensor network representing a holographic-code state with potentially thousands of qubits is computationally intractable; as such, we follow the simplification proposed in~\cite{farrelly_tn_codes,parallel_decoding_tn_codes}. In what follows, we briefly review this technique. 

In the tensor-network decoder formalism, we represent a stabilizer operator as elements in $\mathbb{Z}_{4}$. As an example, we represent the stabilizer $X_{1}Z_{2}Y_{3}Y_{4}X_{5}I_{6}$ for the $\codepar{6,1,3}$ code as the vector $[1, 3, 2, 2, 1, 0]$. In a stabilizer code, we naturally have a prescription for defining four logical operators: $\{\mathsf{\bar{I}}, \mathsf{\bar{X}}, \mathsf{\bar{Y}}, \mathsf{\bar{Z}}\} \in \mathsf{L}$. These operators can be used to define \emph{logical equivalence classes}, permitting us to formulate a rank-$n$ tensor as
\begin{equation}
\mathsf{T}(\mathsf{L})_{\alpha_{1}, \cdots , \alpha_{n}} :=
\begin{cases}
1 & \text{if $P_{\alpha_{1}} \otimes \cdots \otimes P_{\alpha_{n}} \in \mathsf{SL}$} \\
0 & \text{otherwise}
\end{cases}~.
\label{eq:tensor_decoding}
\end{equation}
Here, we take $\alpha_{1}, \cdots , \alpha_{n}$ to be the mapping of a given stabilizer or logical operator to $\mathbb{Z}_{4}$, as shown above; $P_{\alpha_{1}} \otimes \cdots \otimes P_{\alpha_{n}}$ represents the Pauli stabilizer formed from the inverse mapping, and $\mathsf{SL}$ is the \emph{logical coset} of $\mathsf{S}$ with respect to $\mathsf{L}$.

As a side remark, the tensors of \cref{eq:tensor_decoding} are not, strictly speaking, isometric. However, they facilitate efficient contraction, provided that the proper index contraction sequence is indicated; as such, the full decoding process which we will describe below exhibits a runtime complexity of $\mathcal{O}(n^{2.37})$ in the best case~\cite{bravyi2014efficient_tn_decoding,le2014powers}. More details on the formation of large tensor networks using the aforementioned scheme can be found in~\cite{farrelly_tn_codes,junyu_msc_thesis}.

The tensor-network decoder is a \emph{maximum-likelihood} (ML) decoder, which is shown to be optimal given the error model. The ML decoder generally works by calculating the most-likely correction needed to return the system to the correct code state, given a \emph{syndrome}. Here, we define a syndrome as the set $\mathbf{s} = [s_{1} \cdots s_{n-k}]$ of $\mathbb{Z}_{2}$ parity results from measuring each stabilizer. An error $\mathsf{E}$ is initialized as a probability vector corresponding to each individual physical qubit; this vector is then contracted with the holographic tensor network, permitting us to calculate the syndrome $\mathbf{s} = \mathsf{HE}$, where $\mathsf{H}$ is the parity-check matrix of the holographic code.

After finding the syndrome, we deduce the \emph{pure error} (or \emph{destabilizer}) $\mathsf{E}_{\text{pure}}$ by taking the \emph{Moore-Penrose inverse} (or pseudoinverse) $\mathsf{H}_{\text{pseudo}}^{\dagger}$ of the parity-check matrix, resulting in the equation
\begin{equation}
\mathsf{E}_{\text{pure}} = \mathsf{H}_{\text{pseudo}}^{\dagger}\mathbf{s}~.
\end{equation}
Once the pure error is obtained, we proceed to calculate the probability that an error with the form of $\mathsf{E}_{\text{pure}}$ has occurred on the $n$ physical qubits; this is done by contracting the tensor network with tensors that parameterizes the probability of errors on each physical qubit on the boundary. For independently distributed errors, even an exact contraction is efficient. From this procedure, we obtain the expression
\begin{equation}
\mathbb{P}(\mathsf{L},\mathbf{s}) := \sum_{S\in \mathsf{S}} \text{Prob}[\mathsf{E}_{\text{pure}}|SL]~,
\end{equation}
where $\text{Prob}[\mathsf{E}_{\text{pure}}|SL]$ signifies the probability of a pure error acting on the coset of stabilizer $S \in \mathsf{S}$ with respect to logical operator $L \in \mathsf{L}$. In practice, the decoder must find $L$ such that $\mathbb{P}(\mathsf{L},\mathbf{s})$ is maximized, i.e. $\argmax\big[\mathbb{P}(\mathsf{L},\mathbf{s})\big]$. As an additional condition, the summation over all logical cosets must be equivalent to unity; that is $\sum_{L \in \mathsf{L}}\sum_{S \in \mathsf{S}} \mathbb{P}(\mathsf{L},\mathbf{s}) = 1$. As before, we refer the reader to~\cite{farrelly_tn_codes,parallel_decoding_tn_codes,junyu_msc_thesis} for further details on calculating the logical success rate. 

\section{Results}\label{section:results}

\subsection{Noise Model and Setup} \label{section:noise_model}

The error model tested in this work consists of the code-capacity error model for Pauli noise. Such an error model takes on the form 

\begin{equation}
    \mathsf{E}(\rho)=(1-p)\rho +p(r_X X\rho X+r_Y Y\rho Y+r_Z Z\rho Z) \ ,
\end{equation}

with the relative error probabilities $r_{X} + r_{Y} + r_{Z}=1$, typically in the regime of $r_{X} = r_{Y}$ with a bias $\eta = \frac{r_{Z}}{r_{X} + r_{Y}} \geq \frac{1}{2}$ (in the case of $Z$-biased noise). When $\eta = \frac{1}{2}$, one recovers the standard depolarizing noise channel. If we move $\eta \mapsto 0$, then we necessarily move towards \emph{2-Pauli} noise; that is, noise channels consisting only of two Pauli operators, such as XZ or XY noise. 

The seed tensors and their associated logical operators and their stabilizers are shown in \cref{table:seed_tensors}; we have shown all of the seed tensors' stabilizers and logical operators studied in this work. Here we have used the notation $XX$ to denote $X \otimes X$ for brevity. Additionally, the final entry for each stabilizer generator (with a bar above) denotes the action on the logical index. 

\begin{table*}
\centering
\resizebox{\textwidth}{!}{%
\begin{tabular}{| c | c | c |} 
\hline
Seed Tensor & Stabilizers & Logical Operators \\ [0.5ex] 
 \hline
HaPPY & $XZZXI\bar{I}$, $IXZZX\bar{I}$, $XIXZZ\bar{I}$, $ZXIXZ\bar{I}$  & $ZZZZZ\bar{Z}, XXXXX\bar{X}$\\ 
Tailored $\codepar{7,1,3}$ & $XZIZXII\bar{I}, IXZIZXI\bar{I}, IIXZIZX\bar{I}, XIIXZIZ\bar{I}, ZXIIXZI\bar{I}, IZXIIXZ\bar{I}, ZIZXIIX\bar{I}$ & $ZZZZZZZ\bar{Z}$, $XXXXXXX\bar{X}$ \\
Steane & $XXIIIXX\bar{I}$, $IXXXIIX\bar{I}$, $IIIXXXX\bar{I}$, $ZZIIIZZ\bar{I}$, $IZZZIIZ\bar{I}$, $IIIZZZZ\bar{I}$ & $ZZZZZZZ\bar{Z}$, $XXXXXXX\bar{X}$ \\ 
$\codepar{6,1,3}$ & $ZIZIII\bar{I}$, $XZYYXI\bar{I}$, $XXXXZI\bar{I}$, $IZZXIX\bar{I}$, $XYXYIZ\bar{I}$ & $XZXZII\bar{X}$, $XYYXII\bar{Z}$\\ 
SCF & $XXIXI\bar{I}$, $IIXXX\bar{I}$, $ZIZZI\bar{I}$, $IZIZZ\bar{I}$ & $XIXII\bar{X}$, $IIZIZ\bar{Z}$ \\
CD-Steane & $XZZIIIX\bar{I}$, $XIZXZII\bar{I}$, $XIIIZZX\bar{I}$, $ZXXIIIZ\bar{I}$,$ZIXZXII\bar{I}$, $ZIIIXXZ\bar{I}$ &  $XZZXZZX\bar{X}$, $ZXXZXXZ\bar{Z}$ \\
[1ex] 
 \hline
\end{tabular}%
}
\caption{Seed tensors and their stabilizers, as well as logical operators for select holographic codes.}
\label{table:seed_tensors}
\end{table*}

Threshold calculations were performed using the tensor-network decoder from~\cite{farrelly_tn_codes,parallel_decoding_tn_codes}. These decoders and others will be made public in an upcoming software package for holographic quantum error correction codes~\cite{fan2024lego_hqec}. However, for the purposes of understanding the current work, we focus on surveying biased-noise resilience for a swath of known holographic codes, and not on a systematic treatment of tensor-network decoding methodology~\cite{bravyi2014efficient_tn_decoding}.

\begin{figure}
\centering
\includegraphics[width=0.85\columnwidth]{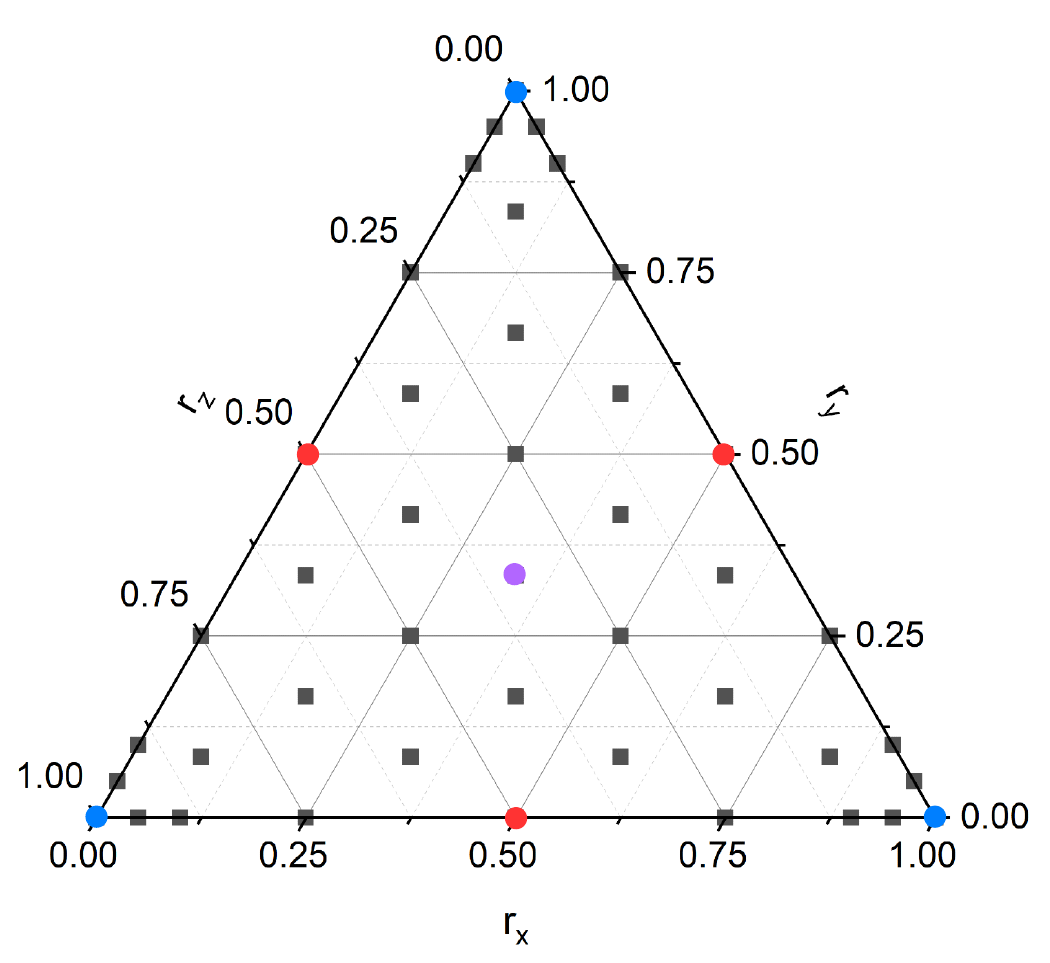}
\caption{Threshold data points surveyed for the holographic codes tested in this work. All data points were ascertained using four threshold crossing points at each corresponding marker in the diagram. In blue, we highlight the pure Pauli biased points, and in red, we have accentuated the 2-Pauli bias points considered in this work. The depolarizing noise point is denoted in violet.}
\label{fig:bias_points}
\end{figure}

In our biased-noise threshold profiling, we tested 43 distinct Pauli biases and have displayed them on the ternary diagram shown in \cref{fig:bias_points}. Each point of the triangular plot shown represents a pure Pauli error channel. Subsequently, points on the interior of the plot can be read by following the grid lines provided to the boundaries; for example, the point to the lower left of the central depolarizing noise marker can be interpreted as having relative biases $\bar{r} = (r_{X}, ~r_{Y}, ~r_{Z})$ of $(0.25, ~0.25, ~0.5)$. At each bias point 10,000 Monte Carlo simulations were performed per threshold curve data point, per layer. We display commensurate threshold curve results for some select data points in \cref{fig:HaPPY_select_threshold_curves} for the zero-rate HaPPY code.

\subsection{Zero-Rate Holographic Code Results} \label{section:zero_rate_results}

\begin{figure*}
\centering
\includegraphics[width=0.9\textwidth]{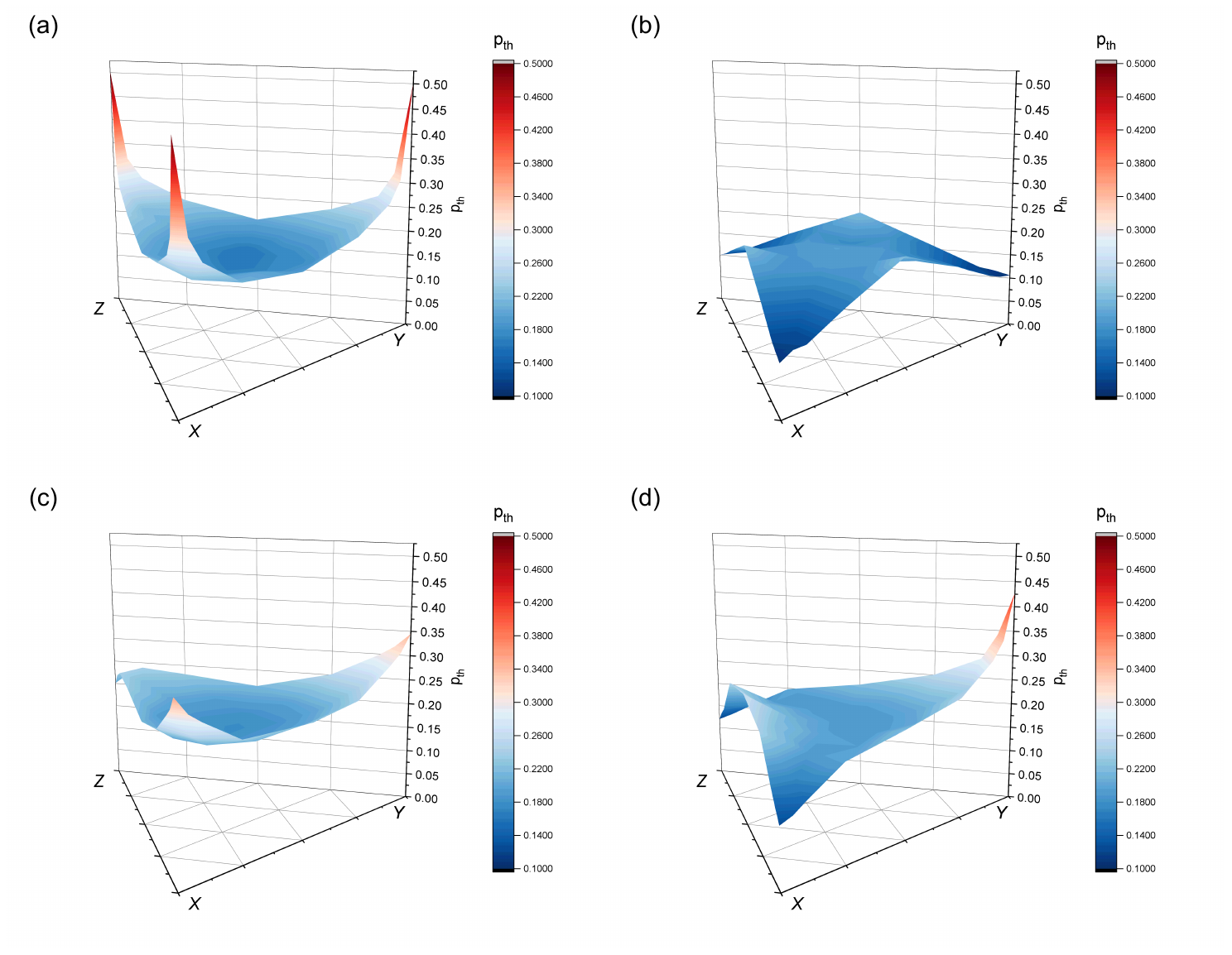}
\caption{Ternary plots for all holographic codes investigated in this study under biased noise, in the code-capacity setting. (a)-(d) depicts the zero-rate HaPPY, Steane, $\codepar{6,1,3}$, and SCF codes; thresholds are color-coded from dark blue ($p_{th} = 10\%$) to dark red ($p_{th} = 50\%$). Additionally, we tested the \emph{tailored} $\codepar{7,1,3}$ code from~\cite{tailored_713,tailored_513_xzzx_codes}, which achieved an identical threshold profile to that of the HaPPY code.}
\label{fig:ternary_plots}
\end{figure*}

\begin{figure*}
\centering
\includegraphics[width=0.9\textwidth]{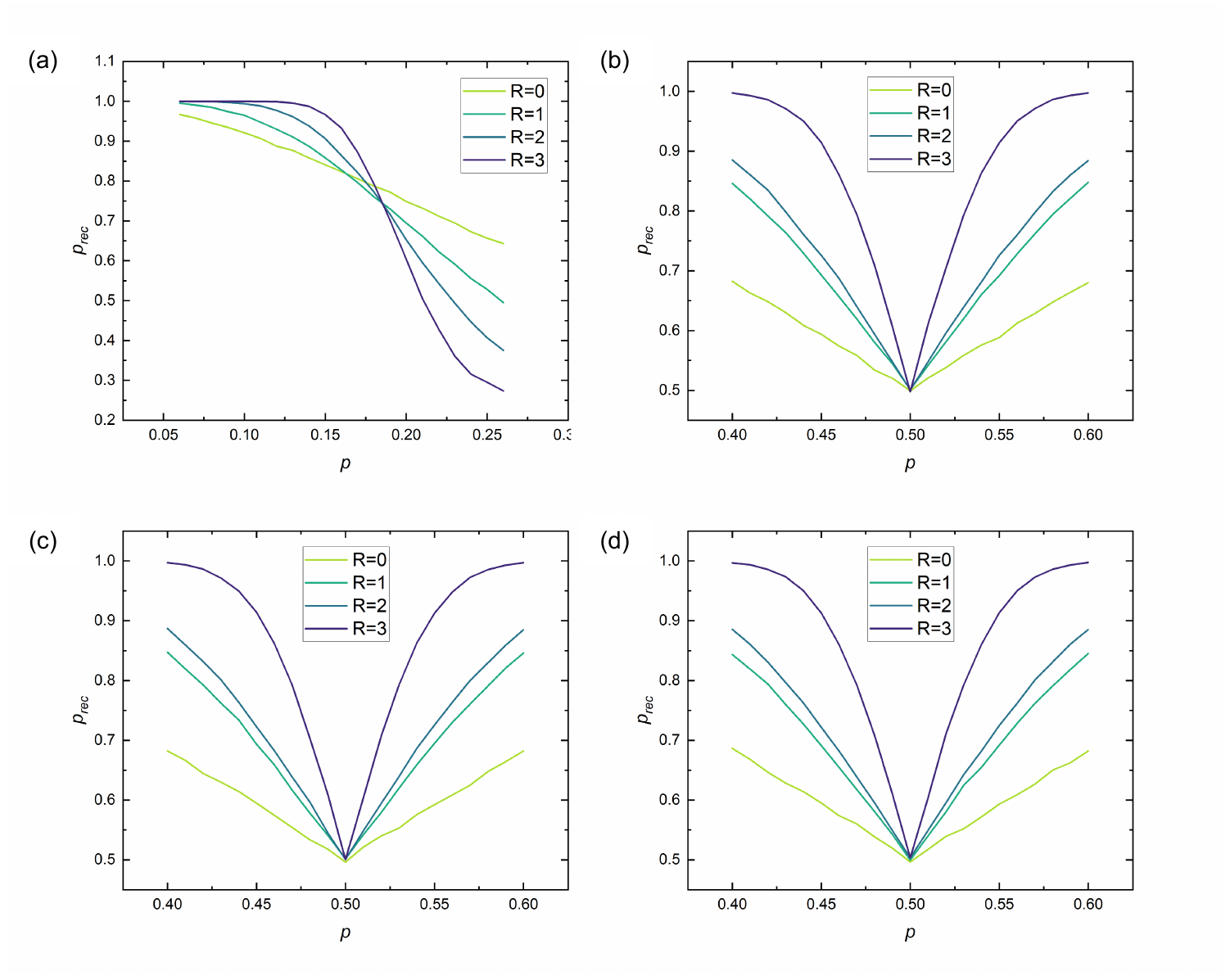}
\caption{Threshold curves for depolarizing, pure $X$, $Y$, and $Z$ noise, as studied using the tensor-network decoder, for the zero-rate HaPPY code at up to $R=3$ layers of edge inflation.}
\label{fig:HaPPY_select_threshold_curves}
\end{figure*}

The ternary plots for all four zero-rate holographic codes tested in this paper are depicted in \cref{fig:ternary_plots}. From (a)-(d), we have displayed the: HaPPY; Steane; $\codepar{6,1,3}$; and SCF. At the center of each diagram, where depolarizing noise is represented, the zero-rate HaPPY code exhibits the lowest resilience ($p_{th} = 17.9\%$), whereas all other codes tested indicating thresholds in the region between $p_{th} = 18\%\sim 19\%$. As we move along each plot towards a pure 1-Pauli bias, we see starkly distinct behavior for every holographic code assessed: for example, the zero-rate HaPPY code (a) attains clear $50\%$ thresholds in each of the pure 1-Pauli biases, while the Steane code's resilience (b) appears to decrease. In the $\codepar{6,1,3}$ and SCF codes, we observe asymmetrical biased-noise profiles with respect to pure Pauli noise behavior. For the $\codepar{6,1,3}$ code, we note that under pure 1-Pauli $X$ and $Y$ noise channels, the threshold of the code increases, albeit more slowly than for the HaPPY code; notwithstanding this similarity, the $\codepar{6,1,3}$ code's threshold dips by a small amount as we move towards the pure-$Z$ region of the ternary plot. Interestingly, the SCF also evinces an asymmetrical biased-noise threshold spectrum. However, the SCF manifests high tolerance to pure $Y$ errors, but to pure $X$ and $Z$ errors decreases, as in the Steane code. 

As a more detailed example, we plotted in \cref{fig:HaPPY_select_threshold_curves} the individual threshold curves obtained for the zero-rate HaPPY code for depolarizing noise, as well as for biased pure $X$, pure $Y$, and pure $Z$ noise. Around the fixed points in subfigures (b)-(d), the recovery rate promptly increases, which is due to several reasons: Firstly, the pure 1-Pauli noise capacity can be shown to be $50\%$. Secondly, as the tensor-network decoder is a maximum-likelihood decoder, after the $p = 0.50$ mark, the decoder selects the most-probable error and returns a pure error (destabilizer) vector which is used to correct. Therefore, unlike a minimum weight decoder, the tensor network decoder is still able to provide the right correction with high probability even when \( p > 0.50 \). More details can be found in~\cite{harris_iod,junyu_msc_thesis}.  

If we look beyond the pure 1-Pauli portion of the ternary plots, we note more subtle behavior of these codes: for example, it can be seen that modest threshold increases emerge for the Steane and $\codepar{6,1,3}$ codes as we move towards pure \emph{2-Pauli} noise, i.e., $XY$, $YZ$, and $XZ$ noise. These changes can be seen more clearly by examining the threshold behavior of the codes via the tuning of a bias parameter $\eta$. In previous work, this aim was accomplished by comparing various zero-rate codes against the \emph{hashing bound}, a useful lower bound for quantifying code-capacity channel performance~\cite{xzzx_surface_code}. However, in contrast to the present work, all previous works have examined the pure 1-Pauli noise limit associated with the hashing bound, and have not delved into combinations of two types of Pauli errors at once, a condition known to occur in the square-lattice GKP code under a symmetric Gaussian random displacement noise channel~\cite{bosonic1,bosonic2,bosonic3,bosonic4}, as well as for error channels resulting from performing \emph{Pauli twirling} on the amplitude damping noise channel~\cite{amp_damping_twirl}. As such, we tune the $\eta$ parameter within the range $0 \leq \eta \leq +\infty$, extending previous work past the regime indicated by $\eta \geq 0.5$~\cite{xzzx_surface_code,clifford_deformed_surface,tailored_513_xzzx_codes}. 

The hashing bound is formally defined as
\begin{equation}
R = 1 - H(\bar{p})~,
\end{equation}

where $R$ represents an achievable rate $k/n$ for a random stabilizer code and $H(\bar{p})$ represents the \emph{Shannon entropy}~\cite{bennett1998quantum} 

\begin{equation}
H(\bar{p}) = -\sum_{i \in \{ I, X, Y, Z \}} p_{i}\log{p_{i}}~.
\end{equation}

Here, $\bar{p} = p\, \bar{r}$, $r_{X} + r_{Y} + r_{Z}=1$ as stated before, and $p$ represents the overall physical error probability. For a given noise model, there exists a physical error probability $p$, at some given relative bias vector $\bar{r}$, for which the achievable rate $R$ goes to zero. This achievable rate via random coding is known as the \emph{zero-rate hashing bound}. 

\begin{figure*}[ht]
\centering
\includegraphics[width=0.9\textwidth]{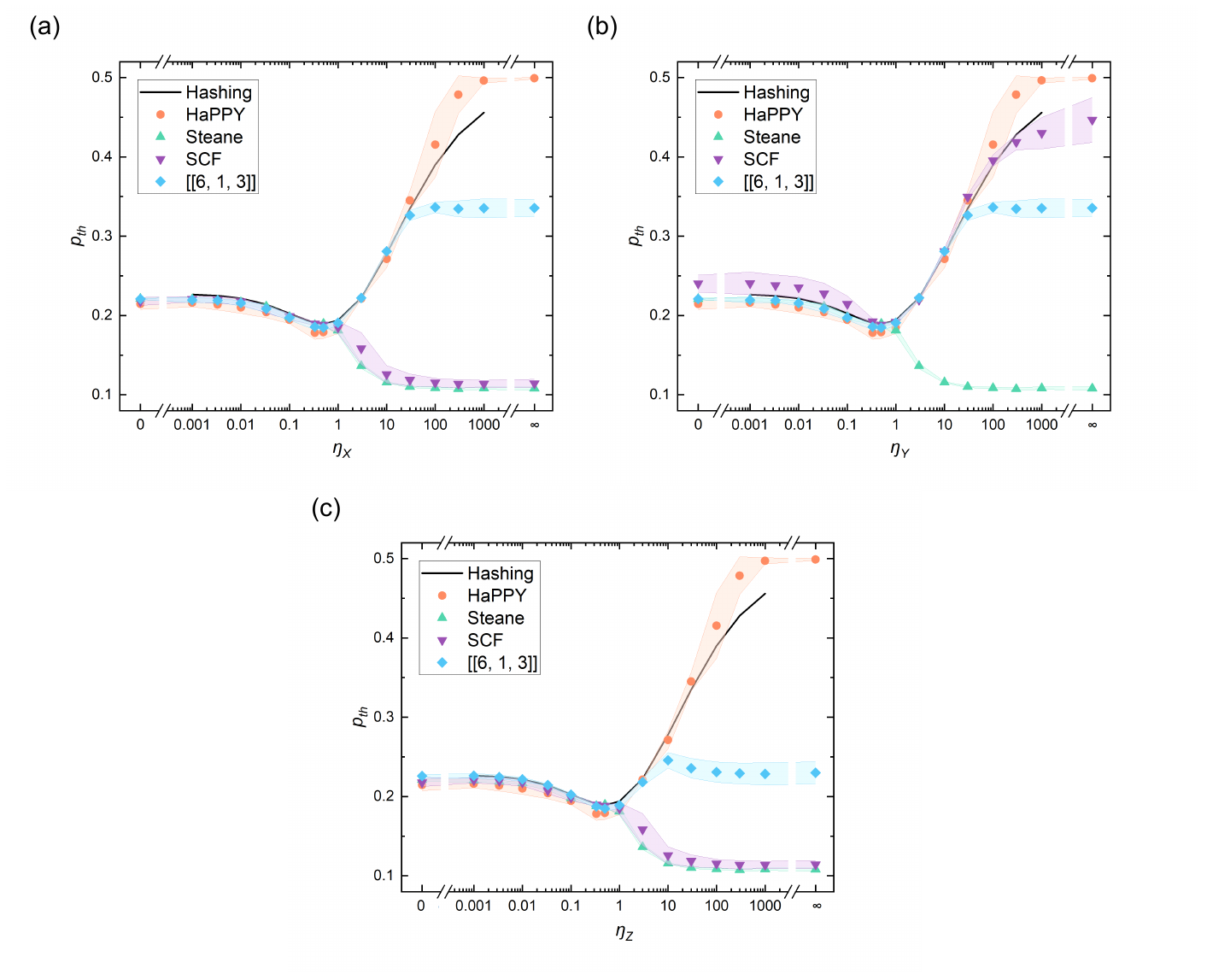}
\caption{Hashing bound plots for all of the codes tested in this work; we denote the hashing bound values in black solid line, while holographic codes are shown in varying colors and markers, together with calculated uncertainties. For all of the plots, biases ranged from $\eta = \{0, +\infty\}$. We list particular points of interest in \cref{table:hb_extremal_points}.}
\label{fig:hashing_bound_plots}
\end{figure*}

\cref{fig:hashing_bound_plots} depicts the comparison of the zero-rate hashing bound for $0 \leq \eta \leq +\infty$ with all four of the surveyed zero-rate holographic codes. In our plots, we calculate physical recovery rates for the following noise biases:

\begin{equation}
\begin{aligned}
    \eta \;\in\; ( & 0, \sfrac{1}{1000}, \sfrac{33}{10000}, \sfrac{99}{10000}, \sfrac{33}{1000}, \sfrac{1}{10}, \sfrac{1}{3}, \sfrac{1}{2},  \\ 
      & 1,3,10,30,100,300,1000,+\infty ) \ .
\end{aligned}    
\end{equation}

In these results, the biased-noise properties of these codes manifest themselves across the full landscape of possible pure and 2-Pauli biases, i.e. pure $X$, $Y$, and $Z$ noise, in addition to pure $XY$, $YZ$, and $XZ$ noise. In (a)-(c), we illustrate the bias tuning for pure $X$, $Y$, and $Z$ biases; consequently, as we tune the parameter $\eta$ in the direction towards zero, we effectively remove $X$, $Y$, and $Z$ noise from the simulations, in effect achieving, as $\eta \to 0$, pure $YZ$ noise from (a), pure $XZ$ noise from (b), and pure $XY$ noise from (c).

\begin{table*}
\centering
{\small  
\begin{tabular}{| c || c | c | c | c || c | c | c |} 
 \hline
Noise bias $(r_{X},r_{Y},r_{Z})$ & HaPPY ($\%$) & Steane ($\%$) & SCF ($\%$) & $\codepar{6,1,3}$ ($\%$) & Hashing ($\%$) & TQEC ($\%$) & Graph Codes ($\%$) \\ [0.5ex] 
 \hline
Depolarizing $(1/3,1/3,1/3)$ & \textcolor{YellowOrange}{$17.9 \pm 0.81$} & \textcolor{ForestGreen}{$18.98 \pm 0.36$} & \textcolor{ForestGreen}{$18.83 \pm 0.13$} & \textcolor{YellowOrange}{$18.46 \pm 0.36$} & $18.929$ & 19.14  & $19.0597$   \\ 
Pure X $(1,0,0)$ & \textcolor{ForestGreen}{$49.92 \pm 0.096$} & $10.79 \pm 0.18$ & $11.42 \pm 0.52$ & $33.56 \pm 1.08$ & $50.00$ & 50.00 & $50.00$\\
Pure Y $(0,1,0)$ & \textcolor{ForestGreen}{$49.91 \pm 0.11$} & $10.8 \pm 0.19$ & \textcolor{YellowOrange}{$44.66 \pm 2.82$} & $33.56 \pm 1.08$ & $50.00$ & $50.00$ & $50.00$\\ 
Pure Z $(0,0,1)$ & \textcolor{ForestGreen}{$49.91 \pm 0.11$} & $10.8 \pm 0.19$ & $11.42 \pm 0.52$ & $22.99 \pm 1.39$ & $50.00$ & $50.00$ & $50.00$\\ 
Pure XZ $(1/2,0,1/2)$ & \textcolor{YellowOrange}{$21.45 \pm 0.68$} & \textcolor{YellowOrange}{$22.1 \pm 0.16$} & \textcolor{RoyalBlue}{$24.027 \pm 1.08$} & \textcolor{YellowOrange}{$22.08 \pm 0.04$} & $22.709$ & {\footnotesize $\approx$Hashing} & $22.684$\\
Pure XY $(1/2,1/2,0)$ & \textcolor{YellowOrange}{$21.45 \pm 0.69$} & \textcolor{YellowOrange}{$22.1 \pm 0.17$} & \textcolor{YellowOrange}{$21.77 \pm 0.51$} & \textcolor{ForestGreen}{$22.58 \pm 0.17$} & $22.709$ & {\footnotesize $\approx$Hashing} & $22.684$\\ 
Pure YZ $(0,1/2,1/2)$ & \textcolor{YellowOrange}{$21.45 \pm 0.68$} & \textcolor{YellowOrange}{$22.1 \pm 0.17$} & \textcolor{YellowOrange}{$21.77 \pm 0.51$} & \textcolor{YellowOrange}{$22.08 \pm 0.04$} & $22.709$ & {\footnotesize $\approx$Hashing} & $22.684$ \\ [1ex] 
\hline
\end{tabular}%
}
\caption{Recovery threshold data points $p_{th}$ for select pure and 2-Pauli biases. The entry in blue surpasses the hashing bound; entries in green attain it up to statistical uncertainty, while entries in gold come within $2\%$ of achieving the bound.The ``HaPPY" column refers to results obtained from both the original HaPPY code of~\cite{happy_paper}, as well as those from concatenating the tailored $\codepar{7,1,3}$~\cite{tailored_713}. In the last two columns, we compare to state-of-the-art results for topological codes (TQEC) as well as for graph codes: Depolarizing-noise thresholds for hexagonal color codes are taken from~\cite{bombin_strong} and pure Pauli noise in surface codes from~\cite{tuckett_biased_surface,xzzx_surface_code}, where thresholds around the Hashing bound were observed near the two-Pauli noise regime (the exact zero-bias regimes were not considered); graph code results are for the 5-in-5 code from~\cite{leditzky_pauli_hashing}.
}
\label{table:hb_extremal_points}
\end{table*}

\cref{table:hb_extremal_points} displays several specific threshold points of interest in the ternary plots from \cref{fig:ternary_plots}. Noise biases and their corresponding bias vectors are displayed, and figures in green and gold signify threshold data which either attain or exceed the hashing bound, or come within $2\%$ of achieving the bound, respectively.

As is evidenced in each of the plots, the zero-rate HaPPY and tailored $\codepar{7,1,3}$ codes clearly achieve the hashing bound for all pure 1-Pauli biases, as well as for finite biases in which $\eta > 10$. Additionally, in the limits for which $\eta \to 0$, the HaPPY and tailored $\codepar{7,1,3}$ code thresholds closely trace out the hashing bound as pure 2-Pauli noise is approached. The behavior of the HaPPY code emulates in large part the behavior seen of the \emph{generalized toric code} family~\cite{tailored_513_xzzx_codes}; we discuss this in more detail in \cref{section:discussion}.

The Steane code portrays a distinct trend: in the pure 1-Pauli noise limit, the code performs significantly worse than all other codes tested. Nonetheless, the zero-rate Steane code eclipses the hashing bound for depolarizing noise ($\eta = 0.5$), while also approaching the bound for finite biases as we approach the pure 2-Pauli limit ($\eta =0$). It was shown in previous work~\cite{tuckett_biased_surface} that the color code exhibits similar behavior in the pure 1-Pauli bias regime; as such, it is logical to infer that the holographic Steane code manifests similar behavior in these limits. Due to the fact that both codes are built from self-dual CSS codes~\cite{lidar_qec}, in the 1-Pauli limit, only one set of stabilizers gives information about the error, while the other stabilizers do not yield additional information for decoding. As such, the problem reduces to a classical linear code with a check matrix defined by only half of the symplectic check matrix in the quantum code. Notwithstanding, it is known that many code concatenation schemes based on graph states can slightly exceed the hashing bound; our results, up to the uncertainty margin given, show that the holographic Steane can at least match most of the codes discovered in~\cite{leditzky_pauli_hashing}.

\begin{figure*}
\centering
\includegraphics[width=0.9\textwidth]{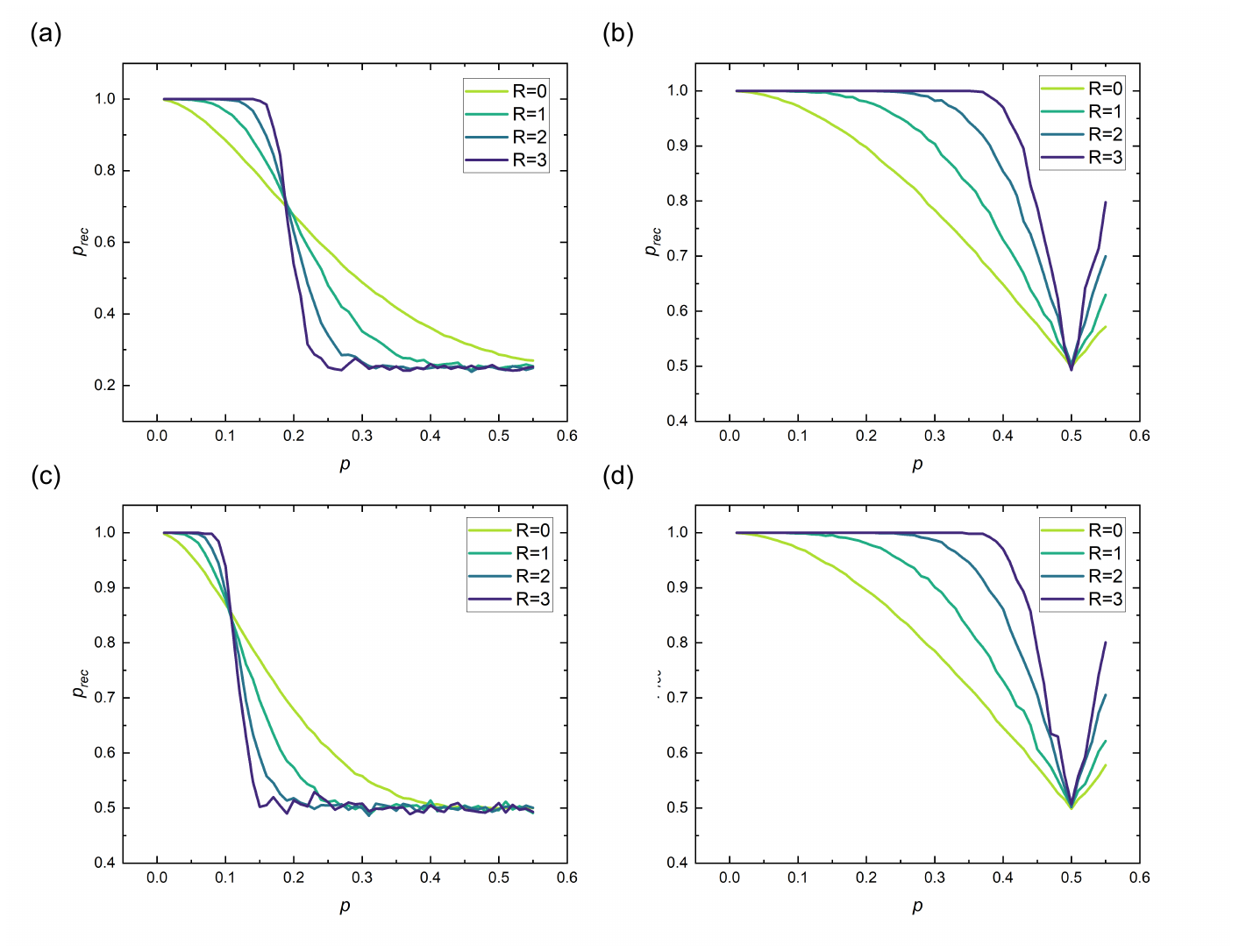}
\caption{Threshold curves for the Clifford-deformed Steane code that we mention in \cref{section:discussion}. Here, we have applied Hadamard gates to physical qubits $2,3,5,$ and $6$. In (a) we document a slight threshold increase to $19.0 \pm 0.33 \%$. (b) and (d) show the thresholds under pure X and Z noise, respectively, under which an increase to approximately $50\%$ can be observed. Finally, (c) displays the results for pure Y noise, with a slight decrease to $10.58\pm 0.12 \%$ apparent.}
\label{fig:clifford_deform_steane}
\end{figure*}

In~\cite{3d_surface_hashing,clifford_deformed_surface}, it was shown that randomized \emph{Clifford deformations}, when applied to the surface code, can improve threshold performance with respect to a code-capacity noise channel. In this way, a surface code may be tailored to achieve the 1-Pauli hashing bound; one wonders whether such randomized approaches could be leveraged in holographic codes as well. Indeed, we may regard the additional Hadamard gates in the Evenbly code~\cite{evenbly_codes2,steinberg_evenbly_codes1} as an instance of Clifford deformation, in order to uphold the strict isometry properties present, while utilizing less highly-entangled quantum states. 

As a preliminary step in this direction, we have performed a random Clifford deformation of the holographic Steane code using Hadamard gates, with the goal of improving the 1-Pauli threshold for the code. The seed tensors used in this modified holographic code are Clifford-equivalent to that of the original Steane code; the only difference lies in the application of Hadamard gates to physical qubits $2,3,5$, and $6$. The resulting stabilizer generators and logical operators for this Clifford-deformed Steane code are given in \cref{table:seed_tensors}. 

As is shown above, the stabilizers and logical operators do not follow the typical convention as expected for mixing Pauli operators in a cyclic manner, nor as in \emph{generalized toric code} constructions~\cite{tailored_513_xzzx_codes,tailored_713,xzzx_surface_code,clifford_deformed_surface}. The thresholds, reported in \cref{fig:clifford_deform_steane}, are for: (a) depolarizing noise ($19.0 \pm 0.33 \%$); (b) pure X noise ($\sim 50\%$); (c) pure Y noise ($10.58\pm 0.12 \%$); and (d) pure Z noise ($\sim 50\%$). We observe that even our naive Clifford deformations, when combined with the holographic concatenation method, can greatly improve the threshold performance for targeting noise models of interest. Indeed, our results show that the Clifford-deformed holographic Steane code achieves the hashing bound for pure X and Z noise, and supersedes the bound for depolarizing noise. Such modifications will be investigated and are the subject of active research.

The SCF displays very interesting behavior at several points in \cref{fig:hashing_bound_plots}. For biased $X$ and $Z$ noise, the SCF performs similarly to the Steane code as $\eta \to +\infty$; this is due to the CSS properties of the seed tensor for both codes, as fewer parity checks are used for distinguishing errors in the limit of pure 1-Pauli noise. For biased $Y$ noise, however, the SCF exhibits excellent threshold performance; although the SCF does not attain the hashing bound in the pure $Y$ noise limit, many points of finite bias reach the bound. Additionally, as we tune the parameter $\eta$ back towards zero, the pure $XZ$ limit shows that the SCF surpasses the hashing bound at finite bias ($\eta \leq 0.1$). To the best of our knowledge, no other code construction thus far has surpassed the hashing bound for 2-Pauli noise~\cite{leditzky_pauli_hashing,lower_bounds_capacity_birgitta}. 

Lastly, the $\codepar{6,1,3}$ code also exhibits breakaway threshold behavior, albeit for pure $XY$ noise. However, it can be observed in \cref{fig:hashing_bound_plots} that various points within the finitely-biased range $\eta \in [0,0.5]$ come either within $2\%$ of the hashing bound, or slightly exceeds it.

\section{Discussion}\label{section:discussion}

In this work we have shown strong evidence of extremely high threshold behavior for zero-rate holographic codes in the presence of 1-Pauli and 2-Pauli noise channels, for pure and finitely-biased noise regimes. In doing so, we have tested the: HaPPY and tailored $\codepar{7,1,3}$ codes; Steane code; $\codepar{6,1,3}$ code; SCF code; and a Clifford-deformed Steane code. All codes have been shown to admit thresholds for some Pauli noise bias which either attain or surpass the hashing bound, and two of these codes, the SCF and the $\codepar{6,1,3}$ code, surpass known the current state of the art for 2-Pauli noise~\cite{leditzky_pauli_hashing,bhalerao2025improving}. Our work demonstrates that holographic codes exhibit remarkable code-capacity properties under biased noise channels and can even overtake current state-of-the-art results~\cite{leditzky_pauli_hashing}. It moreover indicates that they  constitute a novel and competitive class of stabilizer codes that is robust against biased noise, beyond the conventional options of topological codes~\cite{xzzx_surface_code,basudha_xyz2,domain_wall_color_jens,tuckett_biased_surface,3d_surface_hashing,clifford_deformed_surface}, na\"ive constructions of concatenated codes~\cite{exact_performance}, or other advanced techniques~\cite{leditzky_pauli_hashing}. Our results heavily imply that further modification via Clifford deformations is possible, and additionally that holographic codes can in fact be tailored to Pauli noise channels of interest. 


Although our work constitutes the first such study of biased noise and holographic quantum codes, there are still many open directions for future research. Firstly, In related work~\cite{evenbly_codes2}, an integer-optimization decoder was utilized for the zero-rate \emph{Evenbly code} in order to pinpoint depolarizing and pure 1-Pauli noise thresholds of $19.1\%$ and $50\%$, respectively overcoming and achieving the hashing bound. It would be interesting to see if a tensor-network decoder would allow for a more complete biased-noise threshold exploration of Evenbly codes and their many derivatives~\cite{steinberg_evenbly_codes1,evenblyHTN,steinberg_conformal_HTN}. However, in order to realize such an aim, the current tensor-network decoder utilized in our study would need to incorporate the extra Hadamard gates on edges, in order to extract the correct destabilizers from a syndrome. We leave such pursuits for subsequent work. 

Additionally, it is known from~\cite{evenbly_codes2,steinberg_evenbly_codes1} that Evenbly codes exhibit gauge-dependent threshold behavior, owing to the specific isometric constraints used in the structure of the code. In principle, nothing restricts us from utilizing such \emph{gauge-fixing} techniques in conjunction with maximum-rate constructions of the holographic codes studied here. However, the unique seed-tensor isometry properties of each holographic code may be utilized in order to give rise to similar gauge-dependent threshold ``phases"; one example of this can be seen in the maximum-rate HaPPY code under gauge-fixing, which can exhibit very high threshold behavior~\cite{alex_logical_recovery} on a $\{ 5,5 \}$ hyperbolic tiling. A systematic exploration of this subtopic would help to clarify the behavior of holographic codes under gauge-fixing constraints.

Thirdly, regarding the topic of finite-rate holographic code constructions, it is currently unknown whether there exist logical-qubit implantation schemes which maximize a code-capacity threshold for a given error model in a holographic code. It was shown in~\cite{parallel_decoding_tn_codes} that several naive logical-qubit implantation schemes can approach the hashing bound; furthermore, it has been conjectured in~\cite{evenbly_codes2} that certain finite-rate conceptions of Evenbly codes may in fact achieve the code-capacity threshold for quantum erasure noise, as well as the finite-rate hashing bound for Pauli noise. Investigating these possibilities will be at the center of future research efforts.

Recently, generalizations of Abelian stabilizer codes have been investigated as alternatives for low-overhead quantum error correction, as such codes naturally allow for non-Clifford logical gates~\cite{xp_stabilizer,webster2022xp,turaev_viro}. The quantum LEGO formalism itself is not restricted to only stabilizer circuits, and may be readily utilized for simulating aspects of either \emph{non-Abelian stabilizer} or \emph{non-stabilizer} codes. In this case, an enumerator-based method that combines tensor network and Monte Carlo sampling can be used to estimate logical error probabilities and thresholds of the code~\cite{cao2024quantum}. Alternatively, it is possible to calculate the \emph{coherent information} instead~\cite{approx_CFT_codes,coherent_information}. However, new techniques are likely needed to calculate the coherent information for such generalized holographic codes, since the number of physical qubits at the boundary grows exponentially with a commensurate number of layers~\cite{holographic_codes_topical_review,conformal_quasi_holo}.

Comparing all of our results with those of \emph{concatenated codes}, we note that the codes tested here fare better than typical tree-style concatenation~\cite{exact_performance,chamberland2017complementarygauge}, particularly for the depolarizing noise channel. Nevertheless, it was shown in~\cite{leditzky_pauli_hashing} that different types of repetition code concatenations can achieve threshold values comparable to ours for the depolarizing noise limit. Additionally, it was shown that various different concatenation schemes achieve or slightly surpass the hashing bound, depending on the details of the two repetition codes involved in the concatenation. Given this, one may then ask as to why our zero-rate holographic code constructions behave as they do. Considering the structural differences between holographic codes and more straightforward code-concatenation schemes, we would expect that in the 2-Pauli limit a conferrable advantage must be present by concatenation along the edges of a hyperbolic tessellation. Forthcoming work will investigate this advantage on more general footing.

As a final observation from high-energy physics, recent work~\cite{deconfinement} has shown that the bulk geometric transition in AdS/CFT implies a threshold for holographic codes in the continuum limit. Taking this finding in tandem with our results, it is natural to ask whether the \emph{discretized} isometric map typical of holographic codes provides a general method for achieving above-hashing-bound behavior, and additionally, whether or not AdS/CFT in the continuum limit implies the existence of a \emph{fault-tolerance} threshold for holographic codes. 

\subsection{Quantum Coding Theory}\label{section:discussion_quantum_coding_theory}

Here we put our results into perspective for quantum channel capacity and coding theory. Currently, understanding and mitigating the impact of quantum noise is crucial for quantum computation and communication. The most common way to characterize quantum noise is through quantum channels, which provide rigorous descriptions for how a quantum state is transformed or transmitted in the presence of errors~\cite{wilde_qit}; as such, the maximum number of qubits that can be theoretically transmitted reliably in a noisy environment is the \emph{quantum channel capacity} (QCC)~\cite{bennett1998quantum,qccapacity_very_noisy,schumacher1996quantum,schumacher1996sending,privateclassical_qcapacity,lloyd_capacity_noisy}. Determining the precise value of the QCC is therefore known to be incredibly challenging~\cite{Bennett_1997}, and in particular, the QCC is not accurately known even for a channel as ubiquitous as the depolarizing channel. A large part of the difficulty lies in the \emph{superadditivity of coherent information}, which hinders a direct efficient evaluation of the quantum capacity when the channel acts independently and identically over many qubits. As a result, finding lower and upper bounds of the quantum capacity constitute important progress in quantum Shannon theory~\cite{wilde_qit,bennett1998quantum}. 

In this vein, our results stand alongside recent efforts to characterize new classes of quantum codes that come arbitrarily close to the QCC \cite{leditzky_pauli_hashing,xzzx_surface_code}. As the leading method for generating practical QEC codes achieving the QCC relies on random stabilizer codes, our work demonstrates via the holographic code construction that generalizations of code concatenation can achieve the theoretical lower bound for the QCC (i.e., the hashing limit). Such results are of paramount interest, since, as stated above, random quantum stabilizer codes do not generally achieve the optimal transmission rate for their analogous quantum counterparts such as the depolarizing channel or other Pauli channels, unlike their classical counterparts (which are typically capacity-achieving for symmetric channels).

\subsection{Holographic Codes in Practice}\label{section:discussion_holo_codes_in_practice}

We close with a short discussion on the practical implications of our findings. In this work, we have investigated the threshold of holographic codes with respect to \emph{code-capacity} noise channels. Although it has not been confirmed whether or not holographic codes admit a fault-tolerance threshold~\cite{deconfinement}, it seems unreasonable to suspect otherwise, given the fact that the threshold data we have presented here is comparable to many state-of-the-art code families. A first step could be to test \emph{phenomenological} noise models, i.e., those that incorporate measurement noise into the syndrome-extraction process. This goal could in principle be achieved by mapping the tensor network to a \emph{detector picture}, as was recently performed in~\cite{renes_2d}, thereby facilitating large-scale circuit-level noise simulations. 

In the interim, one can consider techniques by which full fault-tolerant syndrome extraction and universal logic can be performed.~\cite{yamasaki1,yamasaki2,goto_many_hypercube} proved that generalized concatenated code schemes based on the quantum Hamming code can exhibit constant space overhead and quasi-polylogarithmic time overhead. These schemes, all based on code concatenation, involve the use of hard-decision layerwise decoding protocols~\cite{knill2005quantum}. As it has been proven that holographic quantum codes are specific instances of generalized concatenated codes for the hyperbolic plane~\cite{heterogeneous_holo_qrm}, we would surmise that the aforementioned techniques can also be adapted to our setting. This idea is the subject of ongoing research. 

Moreover, many alternatives have been suggested as well to the layerwise decoding scheme mentioned above. One method may involve \emph{floquetifying} holographic codes~\cite{hastings2021dynamically}; this procedure would solve two problems at once, in that high-weight stabilizers and logical operators could be controlled to a reasonable weight, and fault-tolerance could be guaranteed with an appropriate measurement schedule. In addition to these ideas, fault-tolerant logic has also been shown to be gauge-dependent~\cite{evenbly_codes2} and universal in some cases for holographic codes~\cite{heterogeneous_holo_qrm}. Recent work has also made progress on lowering the computational complexity of decoding itself using tensor-network schemes~\cite{gray2021hyperkourtis,hyperoptimized_approximate}. These techniques, often based on the analysis of critical or approximate contraction paths, could allow for even faster decoding for error correction and logical operations than is currently employed by tensor-network decoding schemes~\cite{parallel_decoding_tn_codes,farrelly_tn_codes}.

One may also ask what specific applications may be well-suited to the usage of a holographic code. An application of holographic codes may be found in \emph{magic-state distillation}~\cite{magic1,magic2}. Indeed, for the case of the HaPPY and holographic Steane codes, the transversal logical $\overline{SH}$ and $\bar{H}$ gates allow for current state-of-the-art methods to be leveraged, as was pointed out in~\cite{magic3}. For practical quantum error correction, it is known that the square-lattice GKP code admits an induced $XZ$ 2-Pauli noise channel under a symmetric Gaussian random displacement noise model~\cite{bosonic1,bosonic2,bosonic3,bosonic4}, and also that Pauli-twirling the amplitude damping channel results in biased 2-Pauli noise~\cite{amp_damping_twirl}. Such applications could be combined with our work in the future.

Finally, although noise biases can lead to remarkable threshold improvements, it is well known that such gains can be erased in the experimental setting, principally due to complications with gate operations which can unbias the dominant noise channel~\cite{puri2020bias}. As holographic codes approach experimental realization, such considerations will also be necessary.

\subsection{Acknowledgements}

We thank Aritra Sarkar, David Elkouss, and Jens Eisert for discussions related to concatenated codes and quantum channel capacity, and Sivaprasad Omanakuttan for mentioning issues related to bias preservation, principally investigated in~\cite{puri2020bias}. We also thank Mackenzie Shaw and Kathleen Chang for mentioning at the conference FTQC '24 in Benasque several realistic examples of 2-Pauli noise arising in practical quantum computing. MS and SF acknowledge financial support from the Intel corporation.  AJ is supported by the Einstein Research Unit ``Perspectives of a quantum digital transformation''.

\subsection{Author Contributions}

JF developed and implemented the tensor-network decoder during his master thesis, under the supervision of MS and SF. MS, CC, and AJ wrote the manuscript. AJ, CC, and SF provided project guidance.

\clearpage
\bibliography{bibliography}


\end{document}